\documentclass[%
 reprint,
 amsmath,amssymb,
 aps,
]{revtex4-1}

\usepackage{graphicx}
\usepackage{dcolumn}
\usepackage{bm}

\begin{document}

\preprint{APS/123-QED}

\title{Direct measurement of the correlated dynamics of the protein-backbone and proximal waters of hydration in mechanically strained elastin\\}
\author{Cheng Sun, Odingo Mitchell, Jiaxin Huang and Gregory S. Boutis}
 \altaffiliation{Email:gboutis@brooklyn.cuny.edu, Phone: (718)951-5000x2873, Fax: (718)951-4407, Physics Department, Brooklyn College of The City University of New York, Brooklyn NY 11210}

\date{\today}

\begin{abstract}
We report on the direct measurement of the correlation times of the protein backbone carbons and proximal waters of hydration in mechanically strained elastin by nuclear magnetic resonance methods. The experimental data indicate a decrease in the correlation times of the carbonyl carbons as the strain on the biopolymer is increased. These observations are in good agreement with short 4ns molecular dynamics simulations of (VPGVG)$_{3}$, a well studied mimetic peptide of elastin. The experimental results also indicate a reduction in the correlation time of proximal waters of hydration with increasing strain applied to the elastomer. A simple model is suggested that correlates the increase in the motion of proximal waters of hydration to the increase in frequency of libration of the protein backbone that develops with increasing strain. Together, the reduction in the protein entropy accompanied with the increase in entropy of the proximal waters of hydration with increasing strain, support the notion that the source of elasticity is driven by an entropic mechanism arising from the change in entropy of the protein backbone.
\begin{description}
\item[PACS numbers]
87.64.kj, 87.80.Lg
\end{description}
\end{abstract}

\pacs{Valid PACS appear here}
\maketitle

Elastin, the principal protein component of the elastic fiber, is a remarkable biopolymer that gives rise to the resilience of many vertebrate tissues ~\cite{Urry 1988, Tamburro 1999, Keeley 2010}. The structure-function relationship of this system has been of interest since Partridge's original work in isolating elastin from other tissue constituents and a model involving cross-links of hydrophobic and hydrophilic domains ~\cite{Partridge 1955}.  Central to the elasticity of elastin is the degree of protein solvation as well as the polarity of the solvent as made evident by experimental studies of the Young's modulus--elastin only exhibits elasticity when hydrated~\cite{Gosline 1993}. A 'liquid drop elastomer' model was suggested wherein both the surface interaction between hydrophobic and hydrophilic phases and the change in configurational entropy contribute to elasticity ~\cite{Weis-Fogh 1970}. The notion that the elasticity of elastin is entropic in origin was established early on and a model was proposed that incorporated a network of random chains ~\cite{Flory 1974}.  However, experimental work gave evidence of ordered structures in elastin, specifically, the presence of both $\alpha$-helices and $\beta$-turns ~\cite{Urry 1988, Gotte 1968}.  The short model peptide (VPGVG)$_{n}$, an abundant motif of elastin, has been extensively studied as it exhibits the inverse temperature transition that is characteristic of the hydrated protein ~\cite{Urry 1988, Daggett 2001}. In studies of the (VPGVG)$_{n}$ peptide, a model was proposed that incorporates a hydrogen bond formed between Val1 and Val4 residues, resulting in suspended segments between $\beta$-turns~\cite{Urry 2002}. In this model, termed the Librational Entropy Mechanism (LEM), the suspended segments are free to undergo large-amplitude low-frequency torsional oscillations; the amplitude might be damped during extension, giving rise to a decrease in entropy of the segment with increasing strain, providing the driving mechanism for elasticity~\cite{Urry 2002}. In addition, type II $\beta$-turns have also been proposed for the (GXGGX) sequence, a second repeating motif of elastin, wherein hydrogen bonds are formed either between Gly1 and Gly4 residues or between the second and fifth X residues; the interchange between the two conformations results in a dynamical $\beta$-turn sliding giving rise to elasticity ~\cite{Tamburro 1999}. In both model sequences, according to the LEM model, the elastic force is borne by the protein backbone by virtue of a decrease in entropy upon extension. In all elastin models, waters of hydration are tacitly assumed and yet the contribution of proximal waters of hydration to the entropic mechanism of elasticity has eluded experimentalists. Recent short time Molecular Dynamics (MD) simulations of (VPGVG)$_{18}$ have suggested that the orientational entropy of waters hydrating hydrophobic groups decreases, whereas the elastin backbone is more dynamic and has a greater entropy upon extension than in a relaxed state in disagreement with the LEM model~\cite{Daggett 2001}. In this work we report on the first direct measurement of the motional correlation times of proximal waters of hydration and the protein backbone of mechanically strained elastin.

Nuclear Magnetic Resonance (NMR) experiments were performed on purified bovine nuchal ligament elastin purchased from Elastin Products Company, LLC. The elastin was purified by the neutral extraction method of Partridge ~\cite{Partridge 1955}, and was suspended in a mixture of D$_{2}$O and H$_{2}$O with a volume ratio of 50:50 with 0.0003g/mL of sodium azide added as a biocide. The elastin samples were mechanically strained in 1.5cm long, 5mm diameter NMR tubes with both ends held by polytetrafluoroethylene tape and sealed using ethylene-vinyl acetate. Four samples were prepared; the strain ($\Delta$L/L) on the samples was 0, 17, 35 and 43 percent, and are denoted  I, II, III, and IV respectively in the remainder of this work. Additionally, after the experiments were performed on sample IV, the sample was released and measured after 72 hours and is denoted sample V. The loss of water in any of the samples was less than 1 percent over the entire course of experiments. All the experiments were carried out on a 200MHz Tecmag Apollo NMR spectrometer using a Doty solids NMR probe. The 90$^{o}$ pulse time was 9$\mu$s in the $^{13}$C measurements and 19$\mu$s in the $^{2}$H measurements. In the $^{13}$C measurements, a radio frequency field strength of $\omega_{e}$=27.7kHz was applied with $^{1}$H SPINAL-64 decoupling during signal acquisition~\cite{SPINAL64}. All the experiments were conducted at approximately 25$^{o}$C.

We employed the 2D $T_{1}$-$T_{2}$ NMR technique ~\cite{Song 2002}, and measured the $^{2}$H $T_{1}$ and $T_{2}$ relaxation times of water in hydrated elastin. By applying a 2D Inverse Laplace Transform (ILT) of the experimental data, the $T_{1}$ and $T_{2}$ relaxation times are correlated and are manifested into a peak in the 2D map. The 2D ILT map of sample I is shown in Fig. 1 as an example. Four peaks are observed in the figure, corresponding to four reservoirs of water with distinguishable dynamical properties; the 2D ILT maps for all samples studied showed four components. For a spin $I$=1 nucleus, such as $^{2}$H, the relaxation is governed by the quadrupolar interaction and the $T_{1}$ and $T_{2}$ are given by

\begin{equation}
\frac{1}{T_{1}}=\frac{3}{40}(1+\frac{\eta^{2}}{3})C_{q}\{J(\omega_{D})+4J(2\omega_{D})\}
 \label{eq1}
\end{equation} \begin{equation}
\frac{1}{T_{2}}=\frac{1}{80}(1+\frac{\eta^{2}}{3})C_{q}\{9J(0)+15J(\omega_{D})+6J(2\omega_{D})\}
 \label{eq1}
\end{equation}
where $\omega_{D}$ is the Larmor frequency for $^{2}$H, $C_{q}=(\frac{eQ}{\hbar}\frac{\partial^{2}V}{\partial z^{2}})^{2}$ and $\eta$ is the asymmetry parameter of the potential $V$ and assumed to 0 in our case. The spectral density is given by $J(\omega)=\frac{\tau_{c}}{1+(\omega\tau_{c})^{2}}$ with $\tau_{c}$ defined as the correlation time of the fluctuating quadrupolar field of the $^{2}$H nucleus~\cite{Abragam}. Using Eq. 1 and 2 and our measured values of $T_{1}$ and $T_{2}$, the $\tau_{c}$ was determined for the four components shown. Component $\alpha_{1}$ is free water, by virtue of the fact that $T_{1}\approx T_{2}$, $\alpha_{2}$ is water that resides between the elastin fibers and $\beta$ is a mobile component that resides within the elastin fiber as verified by a $T_{2}$-$T_{2}$ exchange experiment ~\cite{Greg 2011}. The water corresponding to the peak with the shortest $T_{1}$ and $T_{2}$ relaxation times, labeled $\gamma$ in Fig. 1, has been determined to be in closest proximity to elastin, also by the $T_{2}$-$T_{2}$ exchange experiment and by considering the correlation times of the slowest two components in the 2D ILT map ($\tau_{c}^{\beta}$=48.8ns and $\tau_{c}^{\gamma}$=111.7ns). In this work, we therefore study the changes in the correlation time of component $\gamma$ as a function of mechanical strain. For samples I, II, III and IV, the correlation times are plotted in Fig. 2a as a function of $\Delta L$/L. It is evident from the figure that the correlation time of proximal waters of hydration decreases with increasing strain on the biopolymer. Not shown in the figure is the correlation time of the water from sample V, which was measured to be the same, within our experimental uncertainty, as that measured in sample I. This observation indicates that the dynamics of the water in close proximity to the protein returned to what was observed in an unstrained sample when the applied strain was removed.

\begin{figure}
 \includegraphics[angle=0,width=4.8cm, height=4cm]{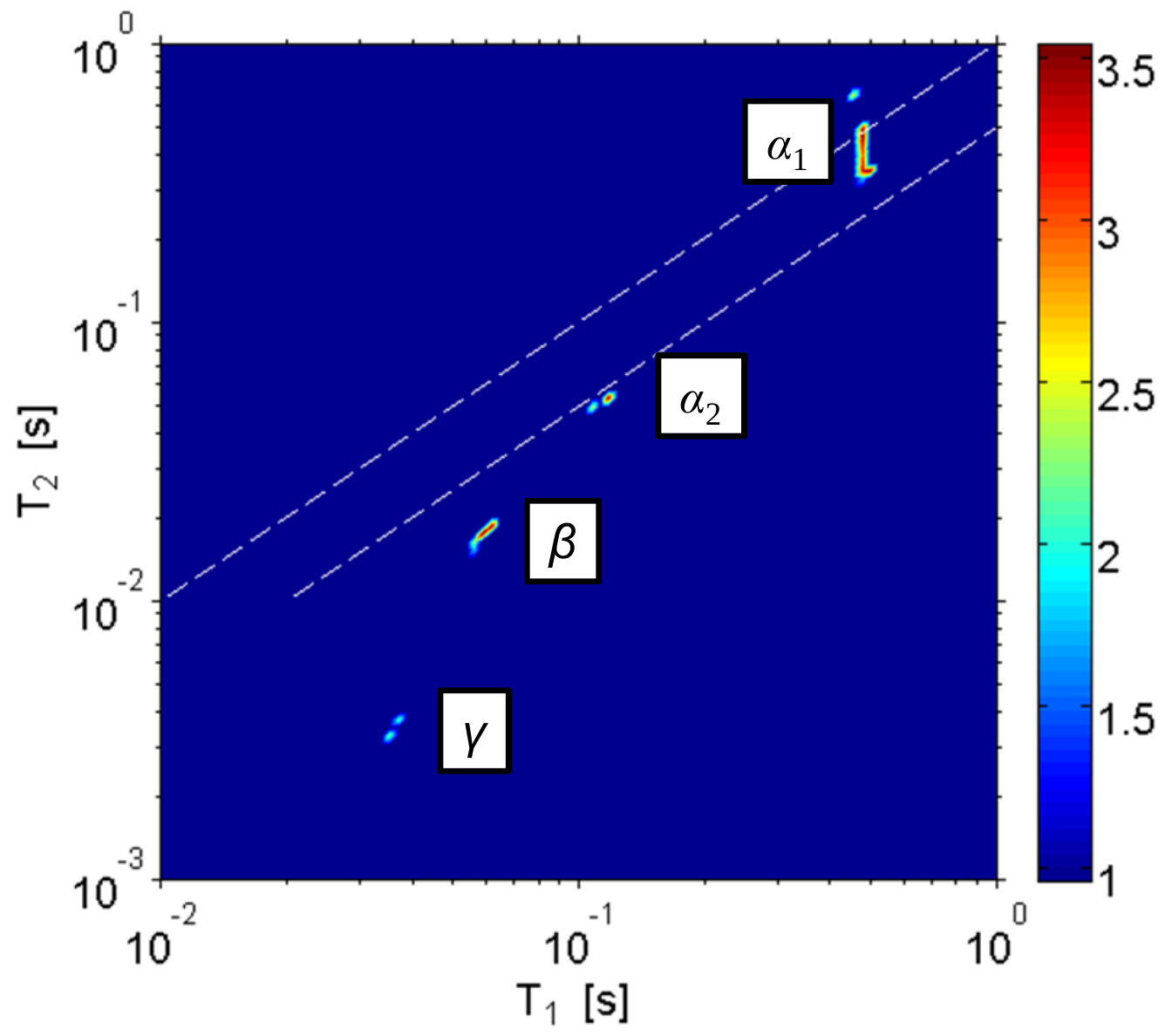} \\
\caption{\label{fig1} $^{2}$H 2D ILT map of the $T_{1}$-$T_{2}$ NMR relaxation times of water in sample I, defined in the text. Four components are discernable and are labeled $\alpha_{1}$, $\alpha_{2}$, $\beta$ and $\gamma$. The dashed lines are used to guide the eye for the region of the 2D map where $T_{1}$ is approximately equal to $T_{2}$. The signal intensity, indicated by the colorbar, is shown on a logarithmic scale.}
\end{figure}

\begin{figure}
 \includegraphics[angle=0,width=7cm,height=4.5cm ]{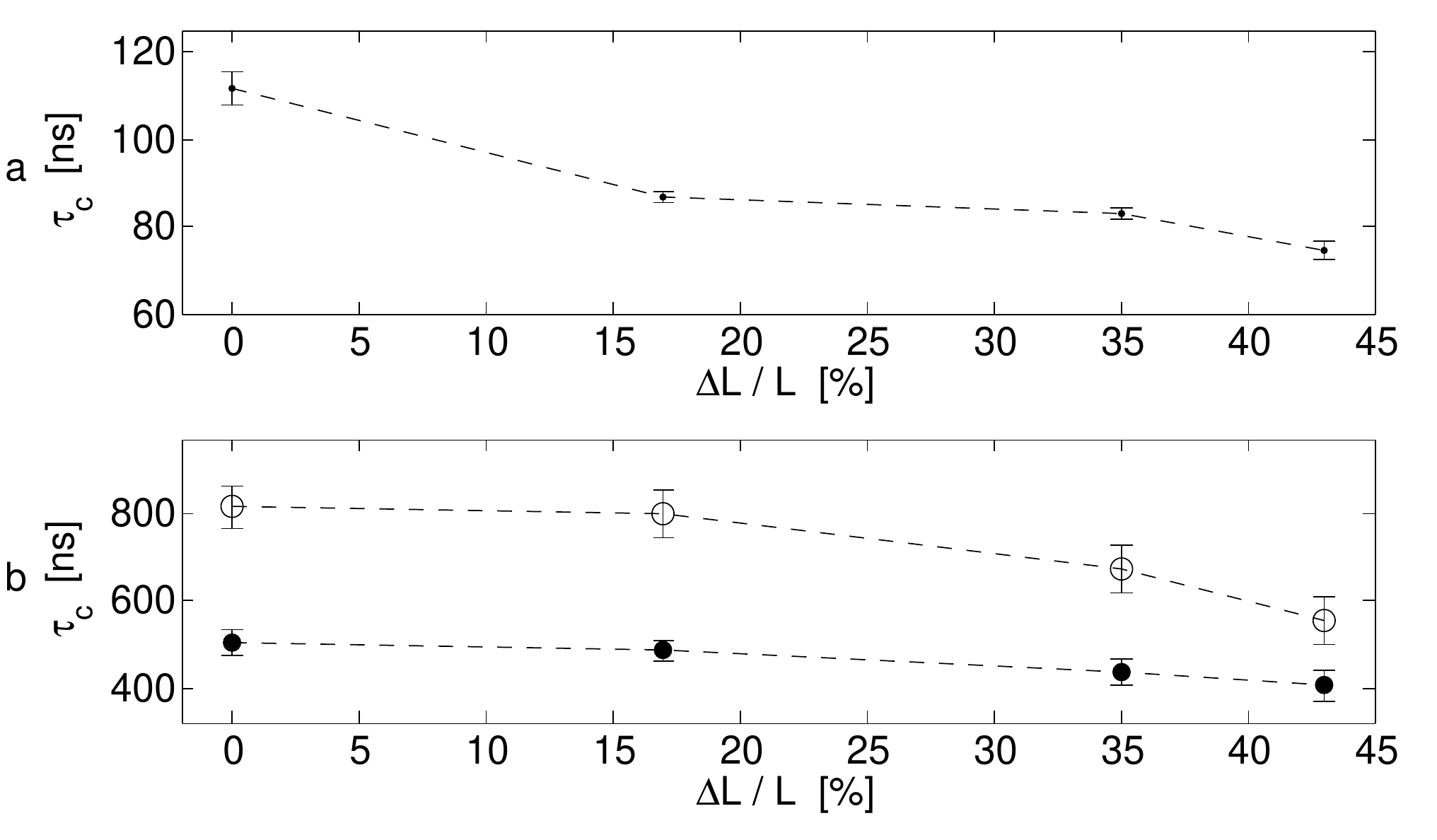} \\
\caption{\label{fig2} Correlation times measured in samples I, II, III and IV as defined in the text. (a) Correlation time of the fluctuating quadrupolar field of D$_{2}$O in close proximity to elastin, as determined from Eq. 1 and 2 and the measured $^{2}$H $T_{1}$ and $T_{2}$. (b) Correlation times characterizing the fluctuating dipolar field experienced by the carbonyl (open circles) and aliphatic carbons (closed circles) of elastin, as determined from Eq. 3 and 4 and the measured $^{13}$C $T_{1}$ and $T_{1\rho}$. The dashed lines are used to guide the eye. }
\end{figure}
Natural abundance $^{13}$C Direct Polarization (DP) and $^{1}$H $\rightarrow$ $^{13}$C Cross Polarization (CP) NMR experiments were performed ~\cite{Duer}, and exemplary spectra are shown in Fig. 3 from sample I. A phase cycling scheme was implemented in the CP experiment, such that the resulting signal observed on the $^{13}$C spectra results only from cross-polarized magnetization. Two features are evident in the spectra. First, it is clear from the DP spectrum, shown in Fig. 3a, that the signal for the carbonyl carbons (centered at 173ppm) is clearly distinguishable from that of the aliphatic carbons (16ppm-60ppm) even without any magic angle spinning~\cite{Kumashiro 2002}; this allows us to distinguish these two signals and measure the respective relaxation times unambiguously.  Second, the overall signal in the CP spectrum (Fig. 3b) is smaller than that of the DP spectrum (Fig. 3a), indicating a high degree of mobility for this system (a factor of 3.977 in gain is expected for cross polarization for the case of a rigid solid ~\cite{Duer}). Similar spectral resolution was achieved on all samples studied.

\begin{figure}
 \includegraphics[angle=0,width=7cm]{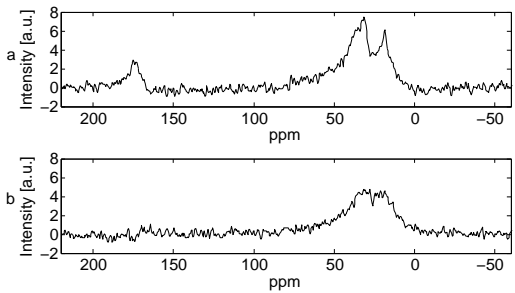} \\
\caption{\label{fig3} Natural abundance $^{13}$C NMR spectra of sample I defined in the text. (a) Direct Polarization (DP) and (b) $^{1}$H $\rightarrow$ $^{13}$C Cross Polarization (CP).  The contact time for $^{1}$H $\rightarrow$ $^{13}$C CP was 1.4ms. The data shown were accumulated with 14512 scans, with a recycle delay of 5s. All spectra are referenced to adamantane.}
\end{figure}

Using the DP technique, we obtained the $T_{1}$ and $T_{1\rho}$ times for the carbonyl and aliphatic carbons of mechanically strained elastin. Given the low spectral resolution of the static sample, one cannot distinguish different signals in the aliphatic region and the dynamics of individual constituents may vary. However, the $^{13}$C $T_{1}$ and $^{1}$H $T_{1\rho}$ values measured on hydrated elastin using magic angle spinning methods at a similar temperature showed similar values for the different components of the aliphatic region ~\cite{Kumashiro 2002}. It is therefore reasonable to study the region by employing an effective $T_{1}$ and $T_{1\rho}$ relaxation time. In hydrated elastin, it has been demonstrated that the carbonyl chemical shift anisotropy is averaged to a great extent due to the high mobility of the system~\cite{Kumashiro 2002}. The $^{13}$C relaxation times are thus mediated largely by carbon-proton dipolar interactions and the $^{13}$C $T_{1}$ and $T_{1\rho}$ are given by
 \begin{equation}
\frac{1}{T_{1}}=\frac{1}{10}C_{D}\{J(\omega_{H}-\omega_{C})+3J(\omega_{C})+6J(\omega_{H}+\omega_{C})\}
 \label{eq3}
\end{equation} \begin{eqnarray}
\frac{1}{T_{1\rho}}=\frac{1}{20}C_{D}\{4J(\omega_{e})+J(\omega_{H}-\omega_{C})+3J(\omega_{C})   \nonumber\\
+6J(\omega_{H})+6J(\omega_{H}+\omega_{C})\}
 \label{eq4}
\end{eqnarray}
where $\omega_{H}$ and $\omega_{C}$ are the Larmor frequencies for $^{1}$H and $^{13}$C respectively, and $C_{D}=(\frac{\hbar^{2}\gamma_{C}\gamma_{H}}{r_{C-H}^{3}})^{2}$~\cite{Wagner 1991}. In the spectral density $J(\omega)$ in Eq. 3 and 4, $\tau_{c}$ is the correlation time of the fluctuating dipolar field experienced by the $^{13}$C spins, which is reflective of the overall motions of the $^{13}$C and nearby $^{1}$H nuclei. Using Eq. 3 and 4, and the measured relaxation times the $\tau_{c}$ is determined for the carbonyl carbons and aliphatic carbons in all the samples. The results are presented in Fig. 2b as a function of the strain on the sample. It is clear from the figure that the correlation times for both carbonyl and the overall aliphatic carbons decrease with increasing strain. The physics of this result dictates an increase in the fluctuation of the dipolar field surrounding both carbonyl and aliphatic regions with increasing strain. The measured correlation times of the carbonyl and overall aliphatic carbons of sample V, not shown, were within the experimental uncertainty of that observed in sample I. To help visualize possible dynamical properties of strained elastin measured in our experimental work, we have studied the elastin mimetic sequence (VPGVG)$_{3}$ by MD simulation.

The MD simulations were performed at 25$^{o}$C using the OPLS-AA/L force field model ~\cite{Tirado-Rives 1988} and a Berendsen thermostat ~\cite{Berendsen} in GROMACS ~\cite{Lindahl 2008}. The peptide was terminated by (-NH-CH$_{3}$) and (-CO--CH$_{3}$) groups at its C- and N-terminus, respectively. The model system was placed in a cubic box solvated with water using the SPC216 water model ~\cite{Engstrom 1987} and was run for 4ns. Two scenarios were studied: (a) without an applied pull force and (b) with the Gly14 atoms being fixed and a pull force being applied on the atoms of Val4 with a constant force equal to 50000 kJ/mol nm. The resulting increase in the radius of gyration, as determined by the position of the C$^{\alpha}$, is highlighted in Table 1 as well as the increase in the number of water molecules in 0.6nm of the peptide resulting from the applied strain.  Further details of the mechanical properties of this elastin mimetic peptide, and others, will appear in a forthcoming manuscript. For each scenario the eigenvalues, $\lambda_{ii}$, of the mass weighted covariance matrix of the MD trajectory were computed. A quasi-harmonic approximation was applied whereby the frequencies, $\omega_{i}$ of uncorrelated simple harmonic oscillators are given by $\omega_{i}=\sqrt{\frac{k_{B}T}{\lambda_{ii}}}$ ~\cite{Karplus 2001}. The entropy of a harmonic oscillator with frequency $\omega_{i}$ at a temperature $T$ is given by
\begin{equation}\label{Eq5}
S_{i}=k_{B}[\frac{\hbar\omega_{i}/k_{B}T}{e^{\hbar\omega_{i}/k_{B}T}-1}-ln(1-e^{-\hbar\omega_{i}/k_{B}T})]
\end{equation}
The histogram of the frequencies derived from the simulations are plotted in Fig. 4 for both scenarios studied. The resulting total entropy of the peptide was determined by summing the entropy contribution of each frequency and is tabulated in Table 1. Fig. 4 shows a shift in population to higher frequency when the peptide is strained. According to Eq. 5 high frequencies contribute less to entropy, thus Fig. 4 implies a decrease in the total peptide entropy when strained; this is observed in our results tabulated in Table 1. In the LEM model~\cite{Urry 2002}, the peptide is treated as a series of quasi-harmonic oscillators undergoing low-frequency, large torsional oscillations when relaxed; the amplitude of the motion is reduced upon extension. Our experimental data, highlighted in Fig. 2b, indicate that there is a decrease in the correlation times of both carbonyl and aliphatic carbons, implying that the motion of the fluctuating dipolar field of both regions increases with increasing strain. We attribute the decrease in the measured correlation times to an increase in frequency of the protein backbone motion as observed in our 4ns MD simulations. The observed increase in backbone motion is also consistent with resulting MD simulations of strained (VPGVG)$_{18}$ ~\cite{Daggett 2001}. In addition, the observed reduction of the amplitude of the backbone motion resulting from the strain, as quantified by the Root Mean Square Fluctuation (RMSF) of the C$^{\alpha}$ shown in Table 1, is also consistent with the expected change according to the LEM model.
\begin{figure}
 \includegraphics[angle=0,width=5cm, height=4cm]{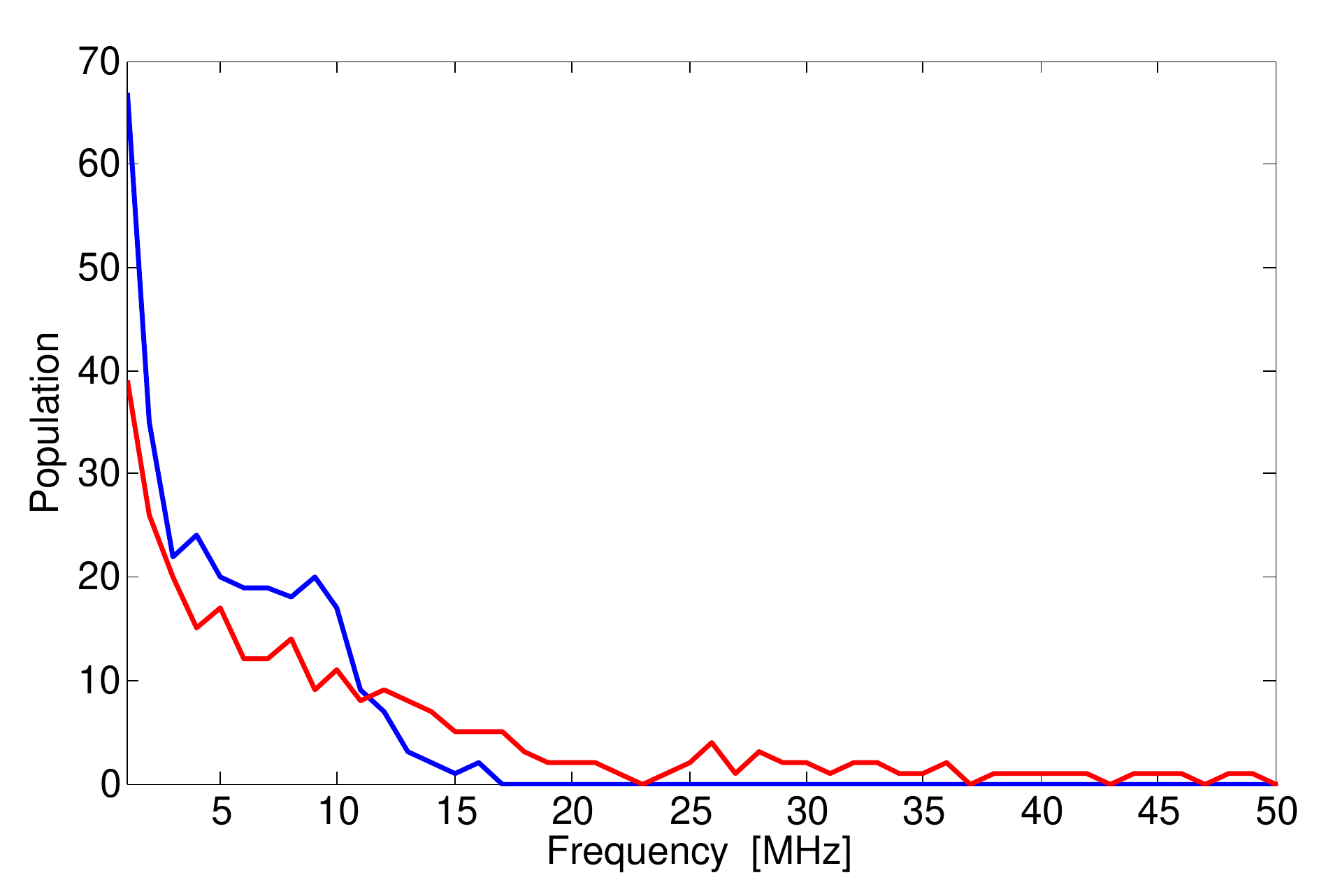} \\
\caption{\label{fig4} Histogram of frequencies derived from the quasi-harmonic approach in the MD simulations of the elastin mimetic peptide (VPGVG)$_{3}$ under relaxed (blue line) and strained (red line) states.}
\end{figure}

\tiny
\begin{table}
\begin{tabular}{|l|l|l|l|l|l|l|l|l|}
  \hline

&  Relaxed&Strained  \\

   \hline
Peptide entropy [kJ/mol$\cdot$K]&1.63&1.46\\
\hline
No. of H$_{2}$O within 0.6nm of peptide& 1132$\pm$248& 5451$\pm$177\\
\hline
Average RMSF of C$^{\alpha}$ [nm]& 0.19$\pm$0.05& 0.14$\pm$0.06\\
\hline
Radius of gyration of C$^{\alpha}$ [nm]&0.78$\pm$0.07&0.83$\pm$0.03\\
\hline
Lifetime of petpide-water H-bond [ps]&0.87$\pm$0.05&0.53$\pm$0.02\\
  \hline

\end{tabular}
\caption{Molecular Dynamics (MD) simulation results for an elastin mimetic peptide (VPGVG)$_{3}$ under relaxed and strained states, as described in the text. Peptide-water H-bonds were counted when the donor-acceptor cutoff distance of 0.35nm and the hydrogen-donor-acceptor angle of 30$^{o}$ was satisfied.}
\end{table}
\normalsize

Referring to Fig. 2a, the measured decrease in correlation time indicates an increase in motion and hence an increase in entropy of the proximal waters of hydration with increasing strain. This observation appears to be correlated with the observation that the lifetime of hydrogen bonds formed between the model peptide (VPGVG)$_{3}$ and water decreases upon extension, as noted in Table 1. Given that the deformation is applied to the protein backbone, it is clear that the changes to the dynamics of the proximal waters of hydration are driven by the increase in the frequency of libration of the backbone. The increase in the entropy of the proximal waters of hydration and reduction in entropy of the backbone upon mechanical strain therefore suggests that the driving mechanism of elasticity is a change in the backbone entropy. While the experimental data suggest that the proximal waters of hydration in elastin increase mobility upon increasing strain, the presence of water is a required, albeit complex element for the elasticity of elastin. Ultimately the solvent gives rise to backbone mobility whose increase in librational frequency with increasing strain reduces the backbone entropy and drives the system back to a relaxed state.

The authors thank Yi-Qiao Song for use of the 2D ILT algorithm and Raymond Tung, Alexej Jerschow, Ranajeet Ghose and Nicolas Giovambattista for useful discussions.  G. S. Boutis acknowledges support from NIH grant number 7SC1GM086268-03. The content is solely the responsibility of the authors and does not necessarily represent the official views of the National Institute of General Medical Sciences or the National Institutes of Health.


\begin{thebibliography}{21}

\bibitem{Urry 1988}
D.W. Urry, J. Protein Chem. $\mathbf{7}$, 1 (1988).
\bibitem{Tamburro 1999}
L. Debelle and A.M. Tamburro, Int. J. Biochem. Cell Biol. $\mathbf{31}$, 261 (1999).
\bibitem{Keeley 2010}
L.D. Muiznieks, A.S. Weiss and F.W. Keeley, Biochem. Cell Biol. $\mathbf{88}$, 239 (2010).
\bibitem{Partridge 1955}
S.M. Partridge, H.F. Davies and G.S. Adair, Biochem. J. $\mathbf{61}$, 11 (1955).
\bibitem{Gosline 1993}
M.A. Lillie and J.M. Gosline, J. Biorheol. $\mathbf{30}$, 229 (1993).
\bibitem{Weis-Fogh 1970}
T. Weis-Fogh and S.O. Anderson, Nature $\mathbf{227}$, 718 (1970).
\bibitem{Flory 1974}
C.A.J. Hoeve and P.J. Flory, Biopolymers $\mathbf{13}$, 677 (1974).
\bibitem{Gotte 1968}
M. Mammi, L. Gotte and G. Pezzin, Nature $\mathbf{225}$, 371 (1968).
\bibitem{Daggett 2001}
B. Li, D.O.V. Alonso, B.J. Bennion and V. Daggett. 2001. J. Am. Chem. Soc. $\mathbf{123}$, 11991 (2001).
\bibitem{Urry 2002}
D.W. Urry and T.M. Parker, J. Muscle Res. Cell Motil. $\mathbf{23}$, 543 (2002).
\bibitem{SPINAL64}
B. M. Fung, A.K. Khitrin and K. Ermolaev, J. Magn. Reson. $\mathbf{142}$, 97 (2000).
\bibitem{Song 2002}
Y.-Q. Song, L. Venkataramanan, M.D. Hurlimann, M. Flaum, P. Frulla and C. Straley,  J. Magn. Reson. $\mathbf{154}$, 261 (2002).
\bibitem{Abragam}
A. Abragam, $Principles$ $of$ $Nuclear$ $Magnentism$ (Oxford University Press, 1961).
\bibitem{Greg 2011}
C. Sun and G.S. Boutis, New J. Phys. $\mathbf{13}$ 025026 (2011).
\bibitem{Duer}
M.J. Duer, $Introduction$ $to$ $Solid-State$ $NMR$ $Spectroscopy$ (Blackwell Publishing Ltd., 2004).

\bibitem{Kumashiro 2002}
A Perry, M.P. Stypa, B.K. Tenn and K.K. Kumashiro, Biophys. J. $\mathbf{82}$, 1086 (2002).
\bibitem{Wagner 1991}
J.W. Peng, V. Thanabal and G. Wagner, J. Magn. Reson. $\mathbf{94}$, 82 (1991).
\bibitem{Tirado-Rives 1988}
W.L. Jorgensen and J. Tirado-Rives, J. Am. Chem. Soc. $\mathbf{110}$, 1657 (1988).
\bibitem{Berendsen}
H.J.C. Berendsen, J.P.M. Postma, A. DiNola and J.R. Haak, J. Chem. Phys. $\mathbf{81}$, 3684 (1984).
\bibitem{Lindahl 2008}
B. Hess, C. Kutnzer, D.V.D. Spoel and E. Lindahl, J. Chem. Theory Comput. $\mathbf{4}$, 435 (2008).
\bibitem{Engstrom 1987}
O. Teleman, B. Jonsson and S. Engstrom, Mol. Phys. $\mathbf{60}$, 193 (1987).
\bibitem{Karplus 2001}
I. Andricioaei and M. Karplus, J. Chem. Phys. $\mathbf{115}$, 6289 (2001).


\end{thebibliography}
\end{document}